\documentclass[aps,twocolumn,superscriptaddress,longbibliography,footinbib,prl]{revtex4-2}

\usepackage[english]{babel}
\usepackage{amssymb,bm,bbold}
\usepackage{mathtools}
\usepackage{graphicx}
\usepackage{multirow}
\usepackage{epstopdf}
\usepackage{latexsym}
\usepackage[normalem]{ulem} 
\usepackage[caption=false]{subfig}
\usepackage[usenames,dvipsnames]{color}
\usepackage{hyperref}
\usepackage{xcolor}
\usepackage{braket}
\usepackage{stackengine}
\usepackage[utf8]{inputenc} 
\DeclarePairedDelimiter\abs{\lvert}{\rvert}%
\DeclarePairedDelimiter\norm{\lVert}{\rVert}%

\begin{document}

\title{Time-reversal in a dipolar quantum many-body spin system
}

\date{\today}

\author{Sebastian Geier}
\affiliation{Physikalisches Institut, Universit\"at Heidelberg, Im Neuenheimer Feld 226, 69120 Heidelberg, Germany}
\author{Adrian Braemer}
\affiliation{Physikalisches Institut, Universit\"at Heidelberg, Im Neuenheimer Feld 226, 69120 Heidelberg, Germany}
\affiliation{Kirchhoff-Institut f\"{u}r Physik, Universit\"{a}t Heidelberg, Im Neuenheimer Feld 227, 69120 Heidelberg, Germany}
\author{Eduard Braun}
\affiliation{Physikalisches Institut, Universit\"at Heidelberg, Im Neuenheimer Feld 226, 69120 Heidelberg, Germany}
\author{Maximilian Müllenbach}
\affiliation{Physikalisches Institut, Universit\"at Heidelberg, Im Neuenheimer Feld 226, 69120 Heidelberg, Germany}
\affiliation{European Center for Quantum Sciences (CESQ-ISIS, UMR7006), University of Strasbourg and CNRS, 23 Rue du Loess, 67200 Strasbourg, France}
\author{Titus Franz}
\affiliation{Physikalisches Institut, Universit\"at Heidelberg, Im Neuenheimer Feld 226, 69120 Heidelberg, Germany}
\author{Martin~G\"{a}rttner}
\affiliation{Physikalisches Institut, Universit\"at Heidelberg, Im Neuenheimer Feld 226, 69120 Heidelberg, Germany}
\affiliation{Kirchhoff-Institut f\"{u}r Physik, Universit\"{a}t Heidelberg, Im Neuenheimer Feld 227, 69120 Heidelberg, Germany}
\affiliation{Institut f\"ur Theoretische Physik, Ruprecht-Karls-Universit\"at Heidelberg, Philosophenweg 16, 69120 Heidelberg, Germany}
\affiliation{Institute of Condensed Matter Theory and Optics,  Friedrich-Schiller-University Jena, Max-Wien-Platz 1, 07743 Jena, Germany}
\author{Gerhard~Z\"urn}
\affiliation{Physikalisches Institut, Universit\"at Heidelberg, Im Neuenheimer Feld 226, 69120 Heidelberg, Germany}
\author{Matthias~Weidem\"uller}
\email{weidemueller@uni-heidelberg.de}
\affiliation{Physikalisches Institut, Universit\"at Heidelberg, Im Neuenheimer Feld 226, 69120 Heidelberg, Germany}

\begin{abstract}

Time reversal in a macroscopic system is contradicting daily experience. It is practically impossible to restore a shattered cup to its original state by just time reversing the microscopic dynamics that led to its breakage. Yet, with the precise control capabilities provided by modern quantum technology, the unitary evolution of a quantum system can be reversed in time. Here, we implement a time-reversal protocol in a dipolar interacting, isolated many-body spin system represented by Rydberg states in an atomic gas. By changing the states encoding the spin, we flip the sign of the interaction Hamiltonian, and demonstrate the reversal of the relaxation dynamics of the magnetization by letting a demagnetized many-body state evolve back-in-time into a magnetized state. We elucidate the role of atomic motion using the concept of a Loschmidt echo. Finally, by combining the approach with Floquet engineering, we demonstrate time reversal for a large family of spin models with different symmetries. Our method of state transfer is applicable across a wide range of quantum simulation platforms and has applications far beyond quantum many-body physics, reaching from quantum-enhanced sensing to quantum information scrambling. 
\end{abstract}

\maketitle  

The concept of time reversal has been a subject of fundamental debate with roots tracing back to the late 19th century. Josef Loschmidt argued that the microscopic laws governing particle interactions were inherently time-symmetric and therefore, the 2nd law of thermodynamics is violated by time-reversing entropy-increasing collisions \cite{loschmidt_uber_1876}. In contrast, Boltzmann remarked on the notion of statistical irreversibility, suggesting that while individual particle interactions might be reversible, the macroscopic behavior of systems exhibited an arrow of time. From the era of Boltzmann and Loschmidt, the concept of time reversal has transitioned from theoretical consideration to practical experiments in the laboratory. Utilizing the precise control capabilities over microscopic degrees of freedom provided by modern quantum technologies, it becomes possible to effectively reverse the arrow of time in the unitary evolution of a quantum system by changing the sign of the Hamiltonian. An early example of this technique is spin echo experiments, where the Hamiltonian and thus the dynamics are reversed by effectively flipping the direction of random magnetic fields that individual spins experience \cite{PhysRev.80.580,Carr1954}. This reversal causes an apparent demagnetized spin state to evolve back-in-time into a magnetized state. While spin echo experiments are based on reversing single-particle dynamics, it is also possible to invert the sign of an interacting many-body Hamiltonian, reversing the buildup of correlations and entanglement in complex states. This type of many-body time-reversal has been demonstrated in a few scenarios, including collective systems \cite{colombo_time-reversal-based_2022,linnemann_quantum-enhanced_2016,Widera2008,garttner_measuring_2017,gilmore_quantum-enhanced_2021}, systems with mixed quantum states \cite{wei_exploring_2018,rovny_observation_2018}, and through a digital approach~\cite{,mi_time-crystalline_2022,braumuller_probing_2022}.

Here, we demonstrate the reversal of quantum dynamics, governed by general many-body spin Hamiltonian which can be tuned by Floquet engineering \cite{geier_floquet_2021,scholl_microwave_2022}. We employ an ultracold atomic Rydberg gas, which has been widely recognized as a highly effective platform for simulating isolated quantum systems owing to the well-decoupled environment \cite{Browaeys2020}. The versatility of the platform enables the exploration of pure quantum states in random \cite{Signoles2019, Orioli2018} and controllable spatial geometries \cite{bluvstein_quantum_2022, Bernien2017,de_leseleuc_observation_2019}. The unitary dynamics in this system are described by spin models with long-range interactions, acting as paradigmatic models to describe quantum magnetism~\cite{Browaeys2020,chen_continuous_2023, Orioli2018, Signoles2019}. Our approach is based on reversing the sign of a many-body spin Hamiltonian by applying a state transfer during the evolution, effectively changing the spin encoding in the Hilbert space. We demonstrate the time reversal by reviving the magnetization of an initially magnetized state after having fully relaxed. Experimental contributions to which the time-reversal is sensitive are identified by employing a Loschmidt echo \cite{Goussev:2012}, becoming sensitive to minute modifications of the many-body wavefunction like, e.g. induced by atomic motion. Finally, by combining the time reversal protocol with Floquet engineering \cite{geier_floquet_2021,scholl_microwave_2022}, we extend the reversal of dynamics to general classes of spin Hamiltonians.

The reversal protocol is illustrated in Fig.~1(a). We encode the (pseudo-) spin in two Rydberg states $\ket{\downarrow}_1 = \ket{nS}$ and $\ket{\uparrow}_1 =\ket{nP}$ (red state combination in Fig. 1(a)). The direct dipolar exchange interactions between  $S-$ and $P-$states realize a Heisenberg XX Hamiltonian
\begin{equation}
    H_{\rm{XX}} = \sum_{i,j} J_{ij}  \left(S_{x}^{i}S_{x}^{j} + S_{y}^{i}S_{y}^{j} \right)\quad,
\end{equation}
where $S_{\alpha}^{i}$ ($\alpha \in {x,y,z}$) are spin-1/2 operators and $J_{ij}= 2C_3 (1 - 3\cos^2\theta_{ij})/r_{ij}^3$.  $C_3$ is the dipolar coupling parameter, $\theta_{ij}$ the angle between atom $i$ and $j$ and the quantization axis, and $r_{ij}$ their spatial separation. Applying two consecutive $\pi$-pulses, we coherently transfer the spin state to another pair of Rydberg states $\ket{\downarrow}_1 \rightarrow \ket{\downarrow}_2 = \ket{n'P'}$ and $\ket{\uparrow}_1 \rightarrow \ket{\uparrow}_2 = \ket{n'S'}$ (blue state combination in Fig. 1(a)). This new encoding realizes the same XX Hamiltonian as in Eq. 1.  However, by properly choosing the state combinations, the coupling parameter changes its sign $C_3 \rightarrow -kC_3$ with $k = |C_3^1/C_3^2|$ being the ratio between the coupling parameters for the two spin encoding subspaces (see SM). Therefore, transferring the spin state between the two sets of encoding Rydberg states effectively realizes a time-reversal operation for the XX Hamiltonian. Encoding the spin in a new set of states has previously been applied to a collective spin model realized with rotational states in polar molecules \cite{li_tunable_2023}.

\begin{figure}
		\centering
		\includegraphics[]{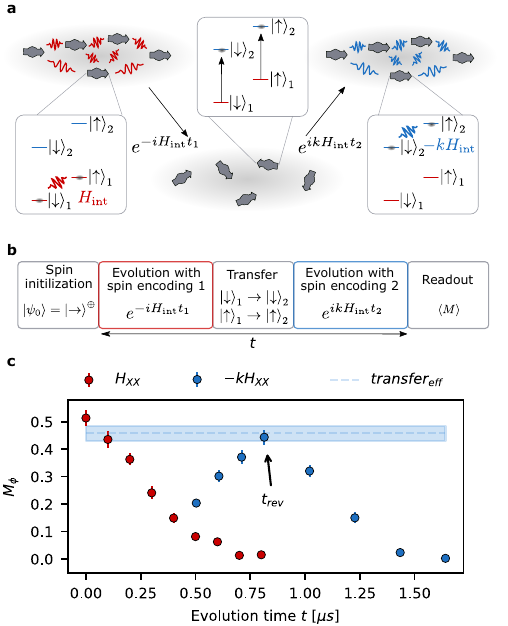}
		\caption{\textbf{Time-reversal on a Rydberg quantum many-body system.} \textbf{(a)} Sketch of the time-reversal protocol. The time-reversal is based on transferring the state between two spin-1/2 encodings in the Rydberg manifold: $\ket{\downarrow}_1= \ket{nS}, \ket{\uparrow}_1= \ket{nP}$ and $\ket{\downarrow}_2= \ket{n'S'}, \ket{\uparrow}_2= \ket{n'P'}$. The unitary evolution in the first spin system is given by $H_{\text{int}}$ and illustrated by the red lines between the spins. Coherently transferring the state into the second spin system will lead to unitary evolution under $-kH_{\text{int}}$, with $k$ being a dimensionless parameter (illustrated by the blue lines between the spins). \textbf{(b)} Experimental protocol to measure time reversal of the magnetization dynamics. The two spin systems are represented by $\ket{\downarrow}_1= \ket{61S_{1/2},m_j = 1/2}$, $\ket{\uparrow}_1= \ket{61P_{1/2},m_j = 1/2}$, $\ket{\downarrow}_2= \ket{61P_{1/2},m_j = -1/2}$, $\ket{\uparrow}_2= \ket{62S_{1/2},m_j = 1/2}$ and the state transfer is performed by two consecutive $\pi-$pulses. \textbf{(c)} Magnetization dynamics in the first spin system without state transfer (red circles) and with state transfer after evolving for $t_1 = 0.4~\mu$s in the first spin system (blue circles). The dashed line displays the state transfer efficiency: the magnetization after transferring the state for $t_1 = t_2 = 0$. The median interaction strength is $J_{\rm{m}}/2\pi = 0.43~$MHz.}
		\label{fig:1}
\end{figure}

\textit{Experimental demonstration.} In our implementation, we employ an ultracold gas of $^{87}$Rb atoms. Notably, this system is spatially disordered due to the random positions of the Rydberg atoms in the cloud \cite{Signoles2019,Franz2022}. The Rydberg density sets a typical length scale which leads to a typical energy scale quantified by the median of the nearest neighbor interaction energy  $J_{\rm{m}} = \mathrm{median}_j \max_i |J_{ij}|$. The experimental sequence demonstrating the reversal of many-body quantum dynamics is shown in Fig. 1(b) and starts with the excitation of Rydberg atoms to $\ket{\downarrow}_1= \ket{61S_{1/2},m_j = 1/2}$ . After the excitation, a microwave $\pi/2-$pulse coupling to  $\ket{\uparrow}_1= \ket{61P_{1/2},m_j = 1/2}$ is used to magnetize the spins along one direction in the equatorial plane of the Bloch sphere. The system then interacts under the XX Hamiltonian with positive $C_3^1/2\pi= 3.2~\rm{GHz}\mu\rm{m}^3$ and evolves into a complex many-body state for a time $t_1$. After the evolution, we coherently transfer the population to the second set of states $\ket{\downarrow}_1 \rightarrow \ket{\downarrow}_2= \ket{61P_{1/2},m_j = -1/2}$ and $\ket{\uparrow}_1 \rightarrow \ket{\uparrow}_2= \ket{62S_{1/2},m_j = 1/2}$, by applying two consecutive $\pi$-pulses with Rabi frequency $\Omega/2\pi = 9~$MHz and $11$~MHz, respectively. For these states, the XX Hamiltonian has a negative interaction coefficient $C_3^2/2\pi= -2.8~\rm{GHz}\mu\rm{m}^3$ and therefore, the ratio between the coupling parameters is $k=1.1$. We let the spins evolve for a time $t_2$. After the second evolution period, the spins are transferred back to the original pseudo spin states where we directly measure the magnetization in the equatorial plane $M_\phi$ using a tomographic readout of the phase contrast \cite{Signoles2019}.

The red circles in Fig. 1(c) show the evolution of the magnetization without state transfer. Starting fully magnetized, the system relaxes towards a demagnetized state within  $\sim0.7~\mu$s, as expected for the underlying XX Hamiltonian \cite{Franz2022}. In contrast, as shown by the blue circles, applying the state transfer after a time  $t_1 = 0.4~\mu$s, where the magnetization has relaxed to $M\sim0.15$,  changes the dynamics drastically. Instead of relaxing further into a fully demagnetized state, the system evolves back into a magnetized state after a time $t_2 = 0.41~\mu$s (or total time $t = 0.81~ \mu s$). This revival of magnetization is expected for a system that evolves back in time or -as done here- for which the sign of the Hamiltonian is reversed as apparent from the unitary evolution of the magnetization $\braket{M_\phi (t_1+t_2)} = \braket{\psi_0 | e^{iH_{\text{XX}} (t_1 - kt_2)} M_\phi e^{-iH_{\text{XX}} (t_1 - kt_2)} | \psi_0}$ which revives to $M_0 = \braket{\psi_0 |  M_\phi| \psi_0} $ at $t_1 = kt_2$. We observe that the obtained value of the coupling ratios $k = 1.03$ is slightly lower than the theoretically expected value of $k = 1.1$. We attribute this difference to interactions being present during the finite-width microwave pulses which are not included in the theoretical value. The light blue dashed line in Fig.~1(c) shows the transfer efficiency, defined as the magnetization obtained after the state transfers with no evolution time i.e. $t_1 = t_2 = 0$, and provides the maximally possible achievable magnetization after the evolution with the reversed Hamiltonian, accounting for infidelities during the transfer. For longer evolution times, the magnetization starts to relax again and ends up in a demagnetized state after $\sim 1.6 ~\mu s$ as expected for a XX Hamiltonian.

\textit{Reversal efficiency.} Time-reversal protocols are generally extremely sensitive to perturbations or decoherence, owing to the complexity of quantum states resulting from many-body interactions. To assess the influence of perturbations such as, e.g., atomic motion or admixture of other atomic states, we characterize the long-time behavior ($J_\text{m}/2\pi\cdot t>1$) of our protocol by measuring the amount of reversed magnetization at the reversal time $t_{\rm{rev}} = t_1 +k\cdot t_1$ (as illustrated in Fig.1(c)) with respect to different evolution times $t_1$ ranging from $0.1~\mu$s to $3~\mu$s. Additionally, we increase the initial Rydberg excitation time, resulting in denser samples with closer atom-to-atom separations, consequently leading to stronger median interactions $J_{\rm{m}}$.

The median interaction strength $J_{\rm{m}}$ is estimated by simulating the spin distribution from a hard-spheres excitation model where each Rydberg excitation is described by a superatom with a given blockade radius and effective Rabi frequency \cite{Signoles2019}. We note that the excitation model is less accurate for high densities, where the distance between Rydberg atoms is on the order of the blockade radius and the hard-sphere approximation breaks down. However, in order to efficiently perform time-reversal experiments, we prepared rather dense samples with strong interactions, such that the interaction timescales are short compared to decoherence times. The obtained spin distribution serves as an estimate of the typical interaction strengths.

For all interactions, the experimental data shows a decrease in the reversed magnetization with time (see circles in Fig. 2(a)), similar to Loschmidt echos with imperfections \cite{Goussev:2012}. For the weakest interactions $J_{\rm{m}}/2\pi = 0.43~$MHz, shown as blue circles, the magnetization still returns to a value of $M\sim0.2$ even after $t_{\text{rev}} = 6 ~\mu$s. To put that into perspective: The magnetization relaxes to zero after only $t_1\sim0.7~\mu$s for this setup. We observe similar behavior for stronger interactions $J_{\rm{m}}/2\pi = 0.79~$MHz (yellow circles) and $J_{\rm{m}}/2\pi = 0.95~$MHz (green circles) despite the enhanced decrease of the overall reversed magnetization as a function of reversal time. 
\begin{figure}
		\centering
		\includegraphics[]{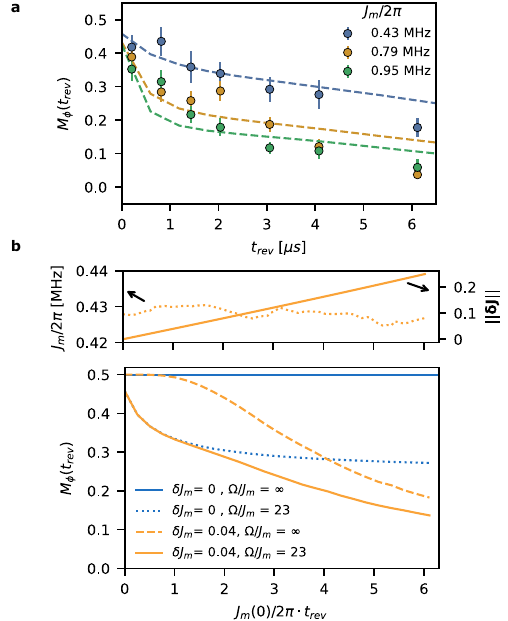}
		\caption{\textbf{Reversal efficiency.} \textbf{(a)} The reversed magnetization measured at the time of the reversal $t_{\rm{rev}} = t_1 +k\cdot t_1$  for various evolution times $t_1$ (circles). The different colors correspond to increasing median interaction strength $J_{\rm{m}}/2\pi$ blue, yellow, and green, respectively. Dashed lines correspond to MACE simulation with a cluster size of 16 atoms.  \textbf{(b)} MACE simulations of the reversed magnetization for $J_{\rm{m}}(0)/2\pi = 0.43$ MHz including different experimental imperfections: No atom motion but finite state transfer pulse width (dotted blue line); atom motion but infinitely fast state transfer (dashed orange line); Atom motion and finite transfer pulse width (solid orange line). The blue solid line shows the situation for perfect reversal with infinitely fast transfer and no atom motion. The upper panel displays the change of the median interaction strength $J_{\rm{m}}$ (dotted orange line) and the matrix norm of $||\delta \mathbf{J}||$ (solid orange line) (see main text for details).}
		\label{fig:2}
\end{figure}	

In an ideal reversal scenario, one would anticipate a complete return to the initial magnetization at all times.
To understand which perturbations influence our time reversal process, we employ simulations of a simplified model, which only considers two internal states $\ket{\downarrow}$ and $\ket{\uparrow}$ per spin. The state transfer pulses are mimicked by a simple $2\pi$ rotation about the $y$-axis at the end of the first evolution time \footnote{We chose the $y$-axis since it does not respect symmetries of the Hamiltonian or the initial state, which the state transfer pulses in the experiment also do not respect.}. The system of hundreds of spins is far too large to solve it exactly, so we perform moving-average cluster-expansion (MACE) simulations~\cite{Hazzard2014}, approximating the dynamics of the full system by simulating small clusters of the $n$ spins and averaging all clusters.

As experimental timescales are more than an order of magnitude shorter than the lifetimes of the included Rydberg states, we neglect these perturbations in our simulations. Instead, we account for two experimental perturbations that might significantly change the system: First, the state transfer Rabi frequencies of $\Omega/2\pi \sim ~10$ MHz are only one magnitude stronger than the median interaction strength in the sample, which will not dominate all interactions considering the maximal interaction strength for particles at the blockade radius can be much larger than $J_\text{m}$. Therefore interactions lead to a modification of the state during the transfer pulses. Second, we include the thermal motion of the atoms during the two evolution periods, which slightly changes the Hamiltonian over time such that a simple sign flip does not reverse the dynamics perfectly. This is done by assigning each atom a fixed velocity $v$ drawn from a Boltzmann distribution at the cloud's temperature $T = 11~ \mu$K and recomputing the couplings every 200 ns according to the changing positions over the course of the simulation. Simulations of this simplified model are shown as dashed lines in Fig. 2(a) and the decrease in reversed magnetization is captured well on a qualitative level. We note, that the cluster sizes of $n = $ 16 atoms are still not providing fully converged simulations (see SM) and therefore our simulation results overestimate the magnetization systematically. This slow convergence in cluster size is attributed to the high sensitivity of Loschmidt echos.

Our simulation allows us to isolate the influence of atomic motion and finite transfer efficiency so we can qualify the effects individually. Focusing on temperature, we study the change in couplings $J_{ij}$ induced by thermal motion. On the one hand, we find the median interaction strength between nearest neighbors $J_\text{m}$ to not vary significantly over the duration of the simulation (dotted line in the upper panel of Fig.~2(b)). This means that the overall distribution of couplings does remain largely the same. On the other hand, directly computing the distance of the full interaction matrix $\mathbf{J}$ as  $\norm{\delta \mathbf{J}} = \frac{\norm{\mathbf{J(t)} -  \mathbf{J(0)}}_F}{\norm{\mathbf{J(t)}}_F}$ (where $\norm{\cdot}_F$ denotes the Frobenius norm) shows a growing deviation very clearly with about $20\%$ difference at 6 interaction cycles (solid line in the upper panel of Fig.~2(b)). We denote the change in couplings over one interaction cycle as $\delta J_\text{m} = \norm{\delta \mathbf{J} (t = 2\pi/J_\text{m})}$. So while the global properties of the coupling distribution do not change, the microscopic configuration does indeed change significantly. The impact on the magnetization can be directly observed by the dashed orange line in the lower panel of Fig.~2(b), displaying the scenario with perfect transfer efficiency $\Omega/J_\text{m} = \infty$ and our finite cloud temperature ($\delta J_\text{m} = 0.04$). The reversed magnetization starts at 0.5 and after an initial plateau starts to continuously decrease over time after a few interaction cycles, due to the sensitivity of the Loschmidt echo to different microscopic configurations. The atomic motion clearly affects the long-term behavior of our time reversal protocol but can not explain the drop in reversed magnetization at short times. 

The exclusive impact of finite state transfer efficiency is investigated at a sample at absolute zero temperature ($\delta J_\text{m} = 0$), so couplings are constant in time but interactions are active during the transfer process (dotted blue line in Fig.~2(b) lower). As expected this reduces the transfer efficiency even at $t=0$ where no dynamics take place except during the transfer pulse. The magnetization reaches a plateau and does not decay further. This means finite pulse times introduce an almost constant error in the magnetization that can be corrected for.

Including both imperfections allows us to qualitatively reproduce our results observed in the experiment (orange solid line). We conclude that for short times ($J_\text{m}/2\pi \cdot t_{\rm{rev}} <$  1), the finite transfer efficiency is the dominating perturbation, while the longer-term behavior ($J_\text{m}/2\pi \cdot t_{\rm{rev}}>$  1) is dominated by the sensitivity to slight changes in the microscopic configuration, due to atomic motion. 

\begin{figure}
    \centering
    \includegraphics[]{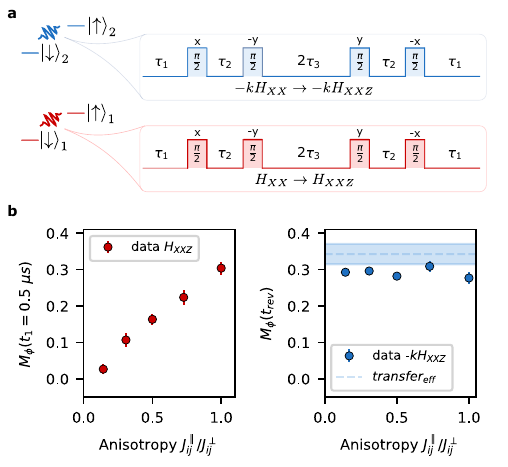}
    \caption{\textbf{Time-reversal of XXZ models with tunable anisotropy.} \textbf{(a)} Protocol: The periodic driving sequence consisting of $\pi/2-$pulses with tunable delay times is applied to the two systems and transforms the natural XX Hamiltonian into an XXZ Hamiltonian with the respective sign. \textbf{(b)} Red circles in the left panel: Magnetization at $t_1 = 0.5~\mu$s as a function of the anisotropy $J_{ij}^\parallel/J_{ij}^\perp$ in the XXZ Hamiltonian without state transfer (for the first spin encoding). Blue points in the right panel: Magnetization at the reversal time $t_{\rm{rev}}$ (for $t_1 = 0.5~\mu$s) after evolving with the same pulse sequence and therefore target Hamiltonian in both spin encoding subspaces, except for the opposite sign. }
		\label{fig:3}
\end{figure}

\textit{Time-reversal of tunable XXZ models.} For quantum engineering applications, it is important to be able to tune the specific type of Hamiltonian one is interested in studying. Quantum metrology applications profit from a power-law XXZ Hamiltonian \cite{perlin_spin_2020} which can directly be combined with time-reversal protocols to enhance phase sensitivity \cite{colombo_time-reversal-based_2022}. On the other hand, investigating quantum information scrambling, e.g. through out-of-time-order correlators, might show distinct behavior for different types of spin models.

In order to achieve the reversal of a wide range of many-body Hamiltonians, we combine our protocol with Floquet engineering, a technique used in various quantum simulators to engineer tunable spin Hamiltonians~\cite{geier_floquet_2021,scholl_microwave_2022,morong_engineering_2023,christakis_probing_2023}. It uses a periodically applied drive to transform a naturally given Hamiltonian into a desired target form. In our implementation, the specific pulse sequence illustrated in Fig.~3(a) transforms the natural XX Hamiltonian into an XXZ Hamiltonian
	\begin{equation}
		H_{\rm{XXZ}} = \sum_{i, j}
		J_{ij}^\perp(S_{x}^{i}S_{x}^{j} + S_{y}^{i}S_{y}^{j})+	J_{ij}^\parallel S_{z}^{i}S_{z}^{j} \quad,
	\end{equation} 
and can be applied to both spin encoding subspaces. Here, $J_{ij}^\perp = J_{ij} \frac{2(\tau_1 + \tau)}{t_c}$ and $J_{ij}^\parallel = J_{ij} \frac{2\tau}{t_c}$, with  $t_c = 2(\tau_1 + 2 \tau)$ being the total sequence time and $\tau_2 = \tau_3 = \tau$ (see \cite{scholl_microwave_2022} for the derivation).  The anisotropy $J_{ij}^\parallel/J_{ij}^\perp$ is tunable with the delay time between the pulses. 

To demonstrate the reversal, we apply a protocol similar to the one introduced in \cite{geier_floquet_2021}. We initialize the system in a state with M$\sim0.4$ magnetization by evolving for $t_{\text{prep}}=100$~ns in the equatorial plane of the spin system with the first spin encoding. We introduce this step to allow the strongest interacting spins in the disordered sample to demagnetize since their interaction cannot be efficiently engineered.  Then, we apply the pulse sequence to engineer XXZ Hamiltonians with anisotropies $J_{ij}^\parallel/J_{ij}^\perp$ between 0.14 and 1 and measure the magnetization at late times $t_1 = 0.5~\mu s$. For $J_{ij}^\parallel/J_{ij}^\perp = 1$ the system possesses a SU(2) symmetry and the magnetization constitutes a conserved quantity while for $J_{ij}^\parallel/J_{ij}^\perp < 1$ we expect a decreasing magnetization due to the breaking of this symmetry. The red circles in the first panel of Fig. 3(b) show the magnetization at $t_1 =$ 0.5 $\mu s$ as a function of the anisotropy. As observed in previous experiments \cite{geier_floquet_2021}, the magnetization increases for increasing $J_{ij}^\parallel/J_{ij}^\perp$ and we almost conserve the full initial magnetization ($M\sim 0.4$) for a value of $J_{ij}^\parallel/J_{ij}^\perp = 1$.
	
After evolution in the first spin state encoding, we transfer the state to the second set of encoding states to flip the sign of the natural Hamiltonian. Applying the very same engineering sequence as in the first half then realizes $-H_{\rm{XXZ}}$ without any other operations necessary. We measure the final magnetization at $t_{\rm{rev}}$, where we expect the revival. The result is shown by the blue points in the second panel of Fig. 3(b). For all probed anisotropies, the reversed magnetization reaches the magnetization expected from the state transfer efficiency, demonstrating the ability to reverse the magnetization dynamics for arbitrary XXZ Hamiltonians.

\textit{Conclusion.} In this work, we demonstrated the time reversal of dynamics, governed by a general tunable  Hamiltonians with power-law interactions, in an isolated quantum many-body system. Our findings lay the foundation for several directions of applications in quantum science. Combined with optical tweezer arrays, the method allows the study of information scrambling dynamics by measuring out-of-time-order correlators \cite{garttner_measuring_2017,braumuller_probing_2022,wei_exploring_2018,swingle_unscrambling_2018} for a more general setting of tunable long-range-interacting spin models. Furthermore, our approach can readily be combined with recent observations of scalable spin squeezing in dipolar systems \cite{bornet_scalable_2023}, potentially enabling phase sensitivity close to the Heisenberg limit \cite{davis_approaching_2016,colombo_time-reversal-based_2022,gilmore_quantum-enhanced_2021,linnemann_quantum-enhanced_2016}.
By time-reversing the evolution of quantum many-body systems, the effect of decoherence can be characterized, constituting an important tool to validate the quality of general platforms exploiting quantum effects, such as entanglement, as a resource~\cite{mi_time-crystalline_2022,rovny_observation_2018}. The time-reversal protocol essentially consists of changing the representation of the pseudo spin in order to realize a flip of the sign of the interaction Hamiltonian and is thus not limited to Rydberg atoms. It can readily be applied to other isolated quantum systems where a plethora of internal states are available. It appears feasible to extend its use to other scenarios, like the Hubbard or the t-J model, where in the latter a state transfer could reverse the sign of the interaction term and the sign of the tunneling term \cite{homeier_antiferromagnetic_2023}. 

\section{Acknowledgments}
\begin{acknowledgments}
We thank C. Hainaut, M. Hornung, D. Lee, V. Salazar, R. Mhaske and A. Bellahasene for helpful discussions. This work is part of and supported by the Deutsche Forschungsgemeinschaft (DFG, German Research Foundation) under Germany’s Excellence Strategy EXC2181/1-390900948 (the Heidelberg STRUCTURES Excellence Cluster), within the Collaborative Research Centre “SFB 1225 (ISOQUANT),” the DFG Priority Program “GiRyd 1929,” the European Union H2020 projects FET Proactive project RySQ (Grant No. 640378), and FET flagship project PASQuanS (Grant No. 817482), and the Heidelberg Center for Quantum Dynamics. The authors acknowledge support by the state of Baden-Württemberg through bwHPC and the German Research Foundation (DFG) through grant no INST 40/575-1 FUGG (JUSTUS 2 cluster) and used the Julia programming for most of the numerics~\cite{bezansonJuliaFreshApproach2017}. T. F. acknowledges funding by a graduate scholarship of the Heidelberg University (LGFG). M. M. acknowledges funding by the European Union H2021 MSCA Doctoral Network QLUSTER (Grant No. 101072964).
\end{acknowledgments}

\newpage
\appendix 
\section{Supplemental Material}
\section{Experimental system and parameter}

Our experiments start with a cloud of ultracold Rubidium 87 atoms in
an optical dipole trap at T = 11 $\mu$K. We optically pump the atoms into the $\ket{g}=\ket{5S_{1/2},F=2,m_F = 2}$ ground state. After turning off the optical dipole trap and excite the $\ket{61S_{1/2},m_j=1/2}$ Rydberg state via  a two-photon laser excitation at wavelengths of 780 nm and 480
nm with a single photon detuning of $\Delta/2\pi$= 97 MHz from the intermediate state $\ket{e}=\ket{5P_{3/2},F=3,m_F = 3}$. During the excitation and the experiment, a B = 78 G magnetic field is applied in order to lift the Zeeman degeneracy in the Rydberg manifold, ensuring a two-level description. The Rydberg excitation time can be varied and changes the Rydberg density (see at the end of this paragraph). To ensure unitary dynamics, we restrict the experimental time to a maximum of 6 $\mu$s, which is short compared to the spontaneous lifetime of the involved Rydberg state (527 $\mu$s for $\ket{61P}$ and 243 $\mu$s for $\ket{61S}$) as well as small compared to the combined lifetime, including blackbody decay (143 $\mu$s for $\ket{61P}$ and 105 $\mu$s for $\ket{61S}$).

The microwave manipulation is realized with an arbitrary waveform generator (Keysight M8195A), directly generating arbitrary waveforms with 16 GHz microwave photons needed to couple e.g. $\ket{61S_{1/2},m_j=1/2}$ to $\ket{61P_{1/2},m_j=1/2}$. After microwave generations, the radiation is focused into our science chamber and drives the spin system. 

The magnetization in the equatorial plane $M_\phi$ is measured at the end of an experimental sequence using a tomographic readout of the phase contrast as detailed in \cite{Signoles2019}. By employing this approach, we account for potential global state detunings, stemming from uncertainties in the magnetic field calibration. These correspond to global rotations about the z-axis, which do not affect the reversal.

Parameter obtained from the excitation model \cite{Signoles2019} for median nearest-neighbor distances $r_{\text{med}}$, blockade radius $r_b$, median interactions strengths $J_{\rm{m}}$, spin excitation time $t_{exc}$, number of spins $N$ are displayed below:

\begin{center}
	\begin{tabular}{ |c|c|c|c|c|c|c|c| } 
		\hline
		&$J_{\rm{m}}/2 \pi$ [MHz] & $r_{\text{med}} [\mu m$] & $r_b  [\mu m$]& $t_{exc} [~\mu s]$ & $N$\\
		\hline
		& 0.43 & 14.3 & 8.1 & 0.8  & 332 \\ 
		& 0.79 & 11.7 & 9   & 1.55 & 917 \\ 
		& 0.95 & 11.1 & 9.6 & 2.55 & 1333 \\ 
		\hline
	\end{tabular}
\end{center}

\section{MACE simulations}

The MACE simulation of the reversed magnetization performed in Fig. 2 of the main text included the finite transfer efficiency and atomic motion. We purely assumed classical thermal motion and neglected motion due to forces between the atoms. This is justified by estimating both effects: 
The distance an atom travels due to the finite cloud temperature is given by $\Delta x = v \cdot t$ with $v = \sqrt{\frac{2k_B T}{m}}$. Therefore, over $1\mu$s the atom moved by roughly $\sim$ 50 nm. Motion due to atomic forces can be estimated by $x = a \cdot t^2$ with $a = F/m = \frac{1}{4\pi\epsilon_0} \frac{3 d^2}{r^4}/m$. Here, $d$ is the dipole matrix element. Over one $\mu$s, the atom moved roughly $\sim$ 0.5 nm.

We note that the MACE simulation of the reversed magnetization performed in Fig. 2 of the main text is not fully converged. However, going to cluster sizes beyond 16 atoms is computationally challenging. Fig. 4 shows the dependence of the reversed magnetization as a function of the cluster size. The slow convergence highlights again the sensitivity of time-reversal protocol with respect to small perturbations.

\begin{figure}
		\centering
		\includegraphics[]{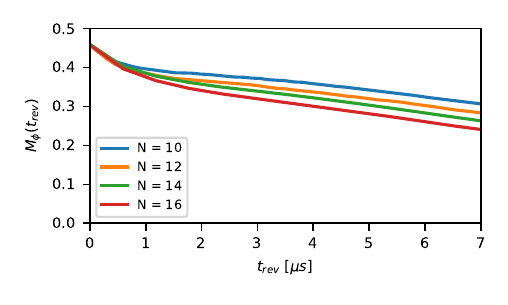}
		\caption{Reversed magnetization as a function of the reversal time for different cluster sizes in the MACE simulations. }
		\label{fig:A1}
\end{figure}

\section{Reversing dipolar interactions}

We motivate our choice of states $\ket{\downarrow}$ and $\ket{\uparrow}$ for the time reversal protocol by considering the explicit form of the interactions. For simplicity, we discuss the case of two atoms whose interactions are described by the dipole-dipole Hamiltonian:
\begin{align}
\hat{H}_{\rm{DDI}} =\frac{1}{4\pi \epsilon_0} \frac{\hat{\mathbf{d}}_1 \cdot \hat{\mathbf{d}}_2 - 3 \left(\hat{\mathbf{d}}_1 \cdot \mathbf{e}_{r}\right)\left(\hat{\mathbf{d}}_2 \cdot \mathbf{e}_{r}\right)}{r^3}\quad,
\label{eq:H_ddi}
\end{align}
where $\hat{\mathbf{d}}_i = (\hat{d}_i^x,\hat{d}_i^y,\hat{d}_i^z) $ is the dipole operator of atom $i$, $\mathbf{e}_{r}$ is the unit vector connecting two atoms and $r$ their distance. For convenience, we change to spherical coordinates such that the dipole operators read 
\begin{align}
	\hat{d}_i^0 &= \hat{d}_i^z \\
	\hat{d}_i^+ &= -1/\sqrt{2} (\hat{d}_i^x + i\hat{d}_i^y) \\
	\hat{d}_i^- &= 1/\sqrt{2} (\hat{d}_i^x - i\hat{d}_i^y)
\end{align}
Here, we chose $z$ as the quantization axis (magnetic field direction). Furthermore, denote $\theta$ as the angle between the atoms and the quantization axis in the following. With this, the dipole-dipole interactions can be expressed as:
\begin{align}
\hat{H}_{\rm{DDI}} = \frac{1}{4\pi \epsilon_0} \Bigg[ \frac{1-3 \cos^2 \theta}{2r^3} \left( 2\hat{d}_1^{\,0}\hat{d}_2^{\,0} + \hat{d}_1^+ \hat{d}_2^- + \hat{d}_1^- \hat{d}_2^+ \right) \label{eq:2_dmj_0} \\
-  \frac{3 \sin \theta \cos \theta}{\sqrt{2}r^3}\left(( \hat{d}_1^+ \hat{d}_2^{\,0}+ \hat{d}_1^{\,0} \hat{d}_2^+)e^{-i \phi} - (\hat{d}_1^{\,0} \hat{d}_2^- + \hat{d}_1^- \hat{d}_2^{\,0}) e^{i \phi}  \right) \label{eq:2_dmj_1} \\
- \frac{3 \sin^2 \theta}{2r^3} \left(  \hat{d}_1^+ \hat{d}_2^+ e^{-2 i \phi} +  \hat{d}_1^- \hat{d}_2^- e^{2i \phi} \right) \Bigg] \quad. \label{eq:2_dmj_2}
\end{align}

In the main text, we only consider interactions that conserve the total angular momentum (this is justified by the application of a magnetic field B = 78 G, shifting unwanted pair states out of resonance) and therefore we are left with the first term in Eq.~\ref{eq:2_dmj_2} providing resonant dipole-dipole interactions
\begin{align}
\hat{H}_{\rm{DDI}} = &\frac{1}{4\pi \epsilon_0} \frac{1-3 \cos^2 \theta}{2r^3} \left( 2\hat{d}_1^0\hat{d}_2^0 + \hat{d}_1^+ \hat{d}_2^- + \hat{d}_1^- \hat{d}_2^+ \right) \\
= & \frac{1-3 \cos^2 \theta}{r^3} 
\begin{pmatrix}
0 & C_3 \\
C_3 & 0
\end{pmatrix} \quad.
\label{eq:H_ddidM0}
\end{align}
The second line expresses the Hamiltonian in the basis $\left\lbrace \ket{\downarrow\uparrow}, \ket{\uparrow\downarrow}\right\rbrace $. The interaction coefficient reads
\begin{align}
C_3 = \frac{1}{4\pi \epsilon_0} \bra{\uparrow\downarrow}\hat{d}_1^0\hat{d}_2^0 + 1/2(\hat{d}_1^+ \hat{d}_2^- + \hat{d}_1^- \hat{d}_2^+)\ket{\downarrow\uparrow} \quad.
\end{align}
We note that for a lifted \(m_J\) degeneracy only one of the three terms is relevant depending on the choice of states. For states with $\Delta m_j = 0$, such as $\ket{\downarrow}_1= \ket{61S_{1/2},m_j = 1/2}$, $\ket{\uparrow}_1= \ket{61P_{1/2},m_j = 1/2}$, we obtain $C_3 = \abs{\bra{\downarrow}\hat{d}^0_1\ket{\uparrow}}^2$. For states with $\Delta m_j = \pm 1$, such as $\ket{\downarrow}_2= \ket{61P_{1/2},m_j = -1/2}$, $\ket{\uparrow}_2= \ket{62S_{1/2},m_j = 1/2}$, we obtain $C_3 = -\abs{\bra{\downarrow}\hat{d}^+_1\ket{\uparrow}}^2$.

\bibliography{Combined}
\end{document}